\documentclass[showpacs,superscriptaddress, preprintnumbers,amsmath,amssymb,twocolumn]{revtex4}

\usepackage{graphicx}
\usepackage{bm}
\include{epsf}

\begin{document}
 \preprint{Prepared for  PRE  }

\title{Two-exponent  Lavalette function. \\A generalization  for  
the   case  of adherents to a religious movement}

\author{Marcel Ausloos $^{1,2,}$  }
\affiliation{  associated with eHumanities group,
Royal Netherlands Academy of Arts and Sciences (KNAW), \\Joan Muyskenweg 25, 1096 CJ Amsterdam, The Netherlands\\  ${^2}$Group of Researchers for Applications of Physics in Economy and Sociology (GRAPES), rue de la Belle Jardiniere, 483/0021, \\ B-4031 Liege, Wallonia-Brussels Federation, Belgium\\
}

\begin{abstract}

The  Lavalette function is generalized to a 2-exponent function in order to represent data looking like a sigmoid  on  semi-log plots. A Mandelbrot trick is suggested for further investigations, if more  fit parameters are needed. The  analyzed data is that of the number of adherents to  the main religions in the XXth century.

\end{abstract}


 \pacs{89.65.-s,
02.50.-r, 05.40.-a , 87.23.Ge,
}
\maketitle
\section{Introduction}\label{introduction}

Recently, there have several studies  of activities related to  society dynamics  based on physics ideas \cite{galam04,galam08,stauffer04}. This is part of an aim of physicists toward applying their concepts and methods in sociological fields \cite{paulo3,galambook,roehner2007driving} Among these are religion/religious movement  aspects. 

Let a few recent papers be mentioned   \cite{hayw99,hashemi,zanette_r,religion1,Herteliu,PicoliMendes08,TAMir} 
 in order  to set a frame and pretend that one aspect among others has not been examined. True the evolution of adherents to religious movements has been studied through the  (statistical) distribution of the number of adherents. However,   one cannot (apparently) distinguish between various analytical forms, neither to say whether the distribution is stable.   There is a report on some "heavy tail"   in \cite{religion1}, but  the  discussion and analysis are  quite reduced to the  shortest arguments for some data  snapshot.

One aim of this paper is to report another (simple) approach, the "least effort method"  \cite{z1} about the rank-size law of  religious movements, through the number of adherents for a set of religions, selected through published data. However, a second aim pertains to the findings suggesting that  the discovered law might be more general than exemplified here, - in fact, as being valid for many cases of  sets measured through the number of members/agents. It is found that a  Lavalette law \cite{Lavalette}, generalized in quite another field \cite{JoI1.07.155Mansilla}, 
  is rather  robust, and represents a fine data envelope,  equivalent to a mean  field approximation of the rank-size relationship asymptotic behavior. Thereby, hinting  that such a law, even possibly further generalized,  as suggested below, can serve to a huge array of data interpretation, models, etc., in many  possible research fields, in physics and outside, - and could be usefully examined on basic principles, - outside the present  aims however.

 In Sect. \ref{sec:data}, the  (religious movement adherent) data is  presented and analyzed along the conventional  rank-size  method,  classically bearing upon Zipf-Mandelbrot-Pareto  (ZMP) ideas. 
    Moreover, introducing a Mandelbrot trick  onto   the  generalized Lavalette function,  is   possible and likely of interest.  Such a generalization and subsequent successful applications seem indicating that  possible "super-generalizations",  see Sect. \ref{sec:hyperLav},  have a wide, $\sim$ universal,  content. In fact, it is readily related to the Feller-Pareto  distribution function (for a chosen set of exponents, see below).  Such a finding suggests some  conclusion  found  in Sect. \ref{sec:conclusion}.

\section{Data and Analysis} \label {sec:data}

Due to the unusual type of  data hereby considered for a "scaling  theory" like analysis, some introduction to the   analyzed WCE data \cite{WCE}  is here given.  
 Indeed, a warning is necessary:  it  is   unclear how much distinction was made in the   WCE surveys concerning denominations and sects so called $adstrated$ to the main religions. Nevertheless, admitting some  flexibility  in the definition of  religion and  ambiguity in that of adhesion, the data are as equally valid as many of those reported about agent behavior in the recent sociophysics literature and examined along statistical mechanics approaches.

The data set is taken from the World Christian Encyclopedia (WCE) \cite{WCE}: it gives information on the number of adherents of the world's main religions and their main denominations (56 religions overall).    I have desaggregated the data for the less general denominations, removing also data on so called  "doubly-affiliated", "disaffiliated",  and "doubly-counted religionists", as listed in \cite{WCE}.   Moreover some data refers to $Atheists$ and  $Nonreligious$ persons.  Thereafter for conciseness, those three sets are also called  ''religions''.

\begin{figure}
\centering
 \includegraphics[height=7.8cm,width=8.8cm]{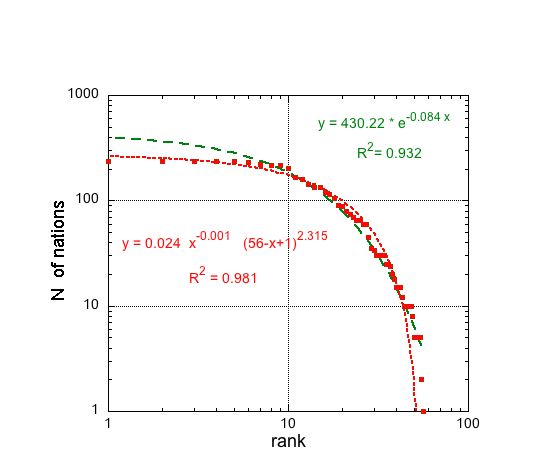} 
\caption{  Cumulative distribution over the XXth  century of the number of nations where a number of religions can be found; the number of countries is ranked in decreasing order of the number of "supported/available" religions}
\label{fig:Plot3lav3Ncountry}
\end{figure}

All religions do not exist in every country. Countries can be ranked according to the number of religions which have local adherents. This leads to a so called Pareto plot for the cumulative distribution of  the (56) religious movements  in the (258) countries: Fig.\ref{fig:Plot3lav3Ncountry}.

On this \underline{log-log  plot},  giving the relationship between the number of nations  ranked according to the number of religions which  can be found locally,  it is obvious that the size-frequency distribution is \underline{surely not} a power law,  - not even shown;   the data has visually a convex shape, and also  hardly looks like a Poisson (exponential)  form (green dash lines)
 the   (2 parameter) exponential case (Exp2)  \begin{equation} \label{EXP2}
 y(r)=  \;b\;e^{-\beta r}.
\end{equation}  
 However,  a fit by  a generalized Lavalette function  (Lav3)  \cite{JoI1.07.155Mansilla} (red dot line), i.e. a decaying power law \underline{with a power law cut-off},
 \begin{equation} \label{Lavalette3}
y(r)=   \Lambda \frac{\big[ r\big]^{-\phi}} { \big[N-r+1\big] ^{-\psi} }
\end{equation}
gives a quite good regression coefficient $R^2 \simeq 0.981$; the exponents have values $-\phi= 0.001$ and $\psi=2.315$  
For completeness,  let it be recalled that  Eq.(\ref{Lavalette3}) is a 2-exponent  (thus 3-parameter) generalization of the  basic  2-parameter  (1-exponent),  (Lav2), introduced by Lavalette \cite{Lavalette}, when discussing impact factor distributions:   
  \begin{eqnarray} \label{Lavrevsigmoidal}
 \nonumber 
y(r)=\kappa\; (N\;r/(N-r+1))^{-\chi}  \\ 
 \; \equiv \; \hat{ \kappa}\;   (r/(N-r+1))^{-\chi}  \\ \nonumber 
 \; \equiv \; \hat{ \kappa}\; \;r^{-\chi}\;(N-r+1)^{+\chi}\; . \end{eqnarray}

\begin{figure}
\centering
 \includegraphics[height=7.8cm,width=8.8cm]{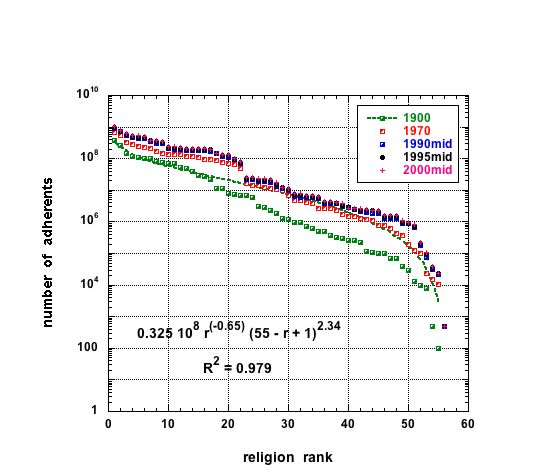} 
\caption{ The \underline{1900 data}  distribution  of adherents to the   55 "major denominations"   (green half open/filled squares) data, with the 3-parameter free Lavalette function,  evolving after $r=18$, toward the recent (after 1970)  data for $r >24$, see Fig. \ref{fig:Plot3religsectfLav3}
}
\label{fig:Plot3religsectf900Lav3}
\end{figure}

In the  WCE  data set \cite{WCE},  further information  is given about  changes in the number of adherents of each (main)  religion from 1900 till 2000, measured over a 5 year span. In Fig.\ref{fig:Plot3religsectf900Lav3}, the \underline{1900 data}  distribution is emphasized:  there were 55 "major denominations", available to  a few billions people. There are nowadays 56 available "major denominations" for about twice as many possible adherents. The 56-th, the "black muslims", came later in the century.  

Observe that a fit of the 1900  (green half open/filled squares) data with the 3-parameter free Lavalette function evolves toward the recent (after 1970)  data for $r >24$.   The fit with a Lavalette 3-parameter free function is very stable as a function of time,  not only matches very well the data for the denominations  with $r>24$, but  also extends toward and covers very finely the "high rank" regime.

  An interesting point,  confirming the warning about data precision, has to be made: it should be emphasized that the  jump in  $N$, the number of adherents, between $r$=24 and 25 is reminiscent of the observation  in \cite{religion1} that it is "somewhat difficult" ({\it 	author's emphasis}) for  major denominations to  measure  their  true number of adherents, - in fact   ({\it daring	author's  interpretation ?}) likely preferring to overestimate such an adhesion in order to appear "big". {\it A contrario}, it seems common sense to consider that  "less popular"  denominations have a better measurement of their number of adherents. The bigger-smaller border  occurs at $\sim 10^7$ adherents, corresponding to $ r\sim 24$.
  
\begin{figure}
\centering
 \includegraphics[height=7.8cm,width=8.8cm]{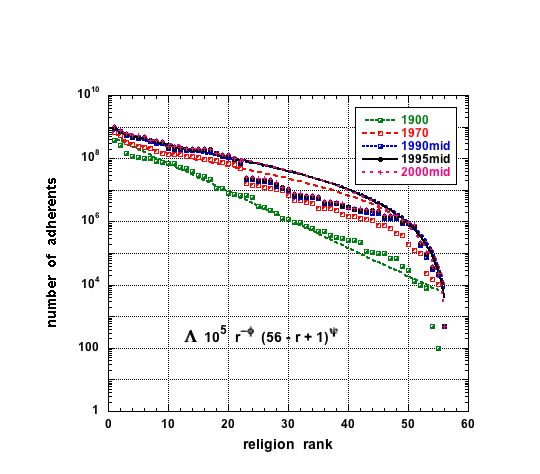} 
\caption{   The more recent data (1970-2000) about the number of adherents to the 56 "major denominations" is analyzed   within the Lavalette "model" (Lav3). Note the slight difference of the 1970 data with respect to the very recent ones, quasi overlapping each other, including the fits of course. A mere exponential fit  ($\sim e^{-0.21\;r}$) is drawn  (green dash line) through the 1900 data  (green half open/filled squares) for visual comparison   
}
\label{fig:Plot3religsectfLav3}
\end{figure}

\begin{figure}
\centering
 \includegraphics[height=7.8cm,width=8.8cm]{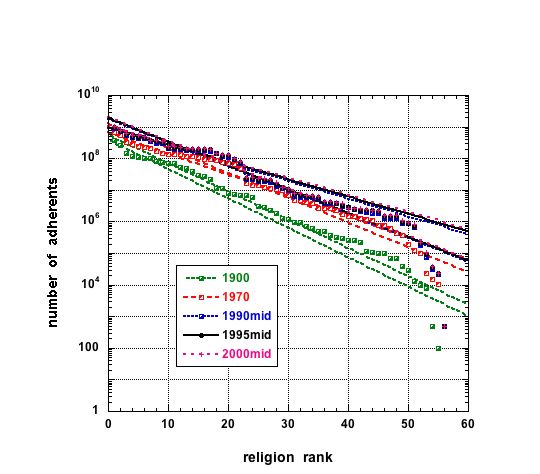} 
\caption{  The whole  data (1900-2000) is analyzed    assuming a mere exponential behavior (Exp2); fit parameters and $R^2$ values are found in Table \ref{Table1} 
}
\label{fig:Plot3religsectfexp}
\end{figure}

  The more recent data (1970-2000) is analyzed in two ways in Figs.\ref{fig:Plot3religsectfLav3}-\ref{fig:Plot3religsectfexp}, either within the Lavalette "model" (Lav3) or within a mere exponential behavior (Exp2). It is more simple to display the parameters in a table, i.e. Table \ref{Table1}, before further discussion. As expected the $R^2$ values are better for the  Lav3 than for the exponential law. Let it be observed that  the $\beta$ decay rate is pretty stable ($\sim0.173$)  in modern times.  The amplitude $b$  follows the evolution of the world population, but  $\Lambda$  decreases, - indicating that people likely become less adherent to  (the main 56) religions, - which can be an "artificial result" due to the fact that the number of small religious movements itself increases, but these are not considered in the WCE surveys.  The power law exponent, $\phi$, measuring the decay, steadily decreases. However the exponent  $\phi$ of the cut-off at high rank  present a minimum (at some time in the middle of the century) and increases nowadays. Note that -($\psi+\phi$) measures the slope at half range on a semi-log plot, and approximately follows the  (increasing nowadays) behavior of $\psi$, i.e. the power law cut-off. This indicates that proportionally the number of adherents to the main religions in fact increases. Another Matthew effect  \cite{Merton} in some sense: the "winning" religions are always  the same ones, and  stay more  at the top than others, even though the number of adherents in the population falls down.

\begin{table} \begin{center} 
\begin{tabular}[t]{cccccccccc} 
  \hline 
   $ $   &1900&1970 &1990&1995&   2000 \\ 

\hline  
   \multicolumn{6}{|c|}{ Generalized Lavalette law (Lav3), Eq.(\ref{Lavalette3})}    \\ \hline  
 $\Lambda\;10^{-5}$ :	& 0.325&	0.684	&	0.325	&0.182	&0.100	\\
  $\phi$ :	& 0.648 &	0.506	&	0.391	&	0.363	&	0.338	 	\\
   $\psi$ :	& 2.341 &	2.289	&	2.554	&2.713	&	2.874	 	\\
  $R^2$ : & 0.979	& 0.984  	& 0.986	& 0.986	& 0.986\\\hline
    \multicolumn{6}{|c|}{Poisson exponential law (Exp2), Eq.(\ref{EXP2})}    \\ \hline  
     $b \;10^{-8}$ :	& 3.995&	6.558	&9.157	&9.771	&10.376	\\
     $\beta $ :	& 0.214&	0.153	&	0.129	&	0.126	&	0.125	\\
  $R^2$ : & 0.937	& 0.945 	& 0.973	& 0.975	& 0.976\\\hline
\end{tabular} 
   \caption{Parameters of  the Lavalette function, Eq.(\ref{Lavrevsigmoidal}), and of the exponential, Eq.(\ref{EXP2}) fits to the number of  adherents to religious movements in the XX-th century according to WCE data   at various years; the number of religions $N_r$ is  55 in 1900, 56 otherwise; the regression coefficient $R^2$  is given}\label{Table1}
\end{center} \end{table}

\section{Hyper-generalized Lavalette function} \label{sec:hyperLav} 

The pure power-law distribution, known as the zeta
distribution, or discrete Pareto distribution is expressed
as
\begin{equation}\label{eq0} p(k) = \frac{k^{-\gamma}}{\zeta(\gamma)}
   \end{equation} 
where:
Ð $k$ is a positive integer usually measuring some variable
of interest,  
Ð $p(k)$ is the probability of observing the value $k$;
Ð $\gamma$  is the power-law exponent;
Ð $\zeta(\gamma)$ $\equiv$$ \sum_{k=1}^{\infty} k^{-\gamma} $ is the Riemann zeta function.
It is important to note, from this definition, that 
$\gamma > 1$
for the Riemann zeta function to be finite. 
Such a mere power law  is of no use here. However,  an  adaptation of  Eq. (\ref{eq0})   was suggested for text appraisal by Mandelbrot, 
 \begin{equation} \label{ZMeq3}
J(r)=b/(\nu+r)^{\zeta},
\end{equation}
in order to take into account a flattening of the data at low rank, on a log-log plot; it is known as the Zipf-Mandelbrot-Pareto 3-parameter free function (ZMP3)  for describing the "queen effect" \cite{Sofia3}. 

Note that
 the role of $r$ as independent variable, in  Eq. (\ref{ZMeq3}),  is taken by the ratio $r/(N - r + 1)$ between the descending and the ascending ranking numbers in Eqs. (\ref{Lavalette3})-(\ref{Lavrevsigmoidal}). Moreover, observe that, on a semi-logarithmic graph, Lav2 and Lav3 follow  a characteristic  (flipped) sigmoidal S-shape which by no means can be provided by a power law or by ZMP3  \cite{JAQMmaLav}. 

Moreover, note that $N$  (as a factor of $r$,  in Eq. (\ref{Lavrevsigmoidal})) is not really needed; in fact it can be usefully  replaced by some simple factor having the order of magnitude of $y(r_M/2)$. This can be seen through
other ways of writing of the 2-parameter Lavalette  form, introducing the scaling factors  $\hat{ \kappa}$ or $\Lambda$.
 
 Finally, if the data fits show such a need, - this is not presently needed, an introduction of Mandelbrot trick into the Lavalette function could be made in order to obtain a  (6-parameter free)  hyper-generalized Lavalette function, i.e.
 \begin{itemize}
\item
  in its most generalized form, with  a power law cut-off
\begin{equation}\label{m1m7Lav}
y(x) = A \;(x+m_3)^{-m_1} \; (N+m_5-x^{m_4})^{m_2}
\end{equation}
 \item
...  or alternatively with an exponential cut-off
\begin{equation}\label{m1m7exp}
y(x) = A \;(x+m_3)^{-m_1} \; e^{-m_2\;(x^{m_4}+m_5)}
\end{equation} \end{itemize}
In the latter formula, $m_5$ is not officially needed: it  can be incorporated into A.
But, beyond such possibilities, it remains  to have some  interpretation of the parameters, thus   much theoretical work  ahead! 

Nevertheless, for concluding this section, observe that the above generalized   in  (2-exponent) Lavalette function can find a root in the Feller-Pareto function:
\begin{eqnarray}\label{FPeq}
\;\;  y(r)=     \Lambda\;  \big[ r\big]^{-\phi}\; \big[N+1-r\big] ^{+\psi}  
  \;  
 \equiv  \hat{\Lambda}\;   u^{-\phi}\;(1-u)^{+\psi}\;,
\end{eqnarray}
when $\phi >0$ and $\psi <0$;  with $u \equiv r/(N+1)$.  To represent a true density function, $\Lambda$ should then be the inverse of a $B $  (Beta) function, for normalization purpose. The Feller-Pareto function is known  to describe random walk distributions,  1-D electronic density of states, moving averages intersections, etc.  However, the Feller-Pareto function has neither the flexibility of the generalized nor the hyper-generalized Lavalette functions \cite{JAQMmaLav}.

\section{Conclusion}  \label{sec:conclusion}

When employing combined
functions,  an excellent fit of rank-frequency distributions is usually  ensured.
However, improving the quality of the approximation at the expense of
an increase in the number of fitting parameters does not provide the best
solution toward inventing an appropriate model. Having a large number of fitting parameters causes difficulties
in the extraction of their implication (meaning) and increases their
dependency on each other,  in worst cases.
The present work, devoted to developing a fitting function that
provides (i) an optimal approximation of rank-probability  
distributions and (ii) a correlation of its fitting parameters with features
of an interesting modern data base, has emphasized that the Lavalette function is of interest in order to represent a sigmoid function with few parameters on  semi-log plots.
 Yet, the basic Lavalette function does not provide enough flexibility for fitting curves in the range of small rank values. Therefore a 2-exponent function has been used, and a Mandelbrot trick suggested for further investigations.

\vskip0.2cm
{\large \bf Acknowledgment}
  The author gratefully acknowledges stimulating and challenging discussions with many wonderful colleagues at  several meetings of the  COST Action MP-0801, 'Physics of Competition and Conflict'.

  \end{document}